\documentclass[runningheads]{llncs}
\usepackage[T1]{fontenc}
\usepackage{graphicx,cite}
\usepackage{float}
\usepackage{mathtools}
\usepackage{amsmath}
\usepackage{amssymb}
\usepackage{color}
\usepackage{hyperref,tabularx}

\begin{document}

\title{Physics-Informed Neural Networks for One-Dimensional Quantum Well Problems}
\titlerunning{Physics-Informed Neural Networks for Quantum Well Problems}
%
\author{Soumyadip Sarkar}
\authorrunning{S. Sarkar}
%
\institute{Department of Computer Application, Narula Institute of Technology, India\\
\email{soumyadipsarkar@outlook.com}}

\maketitle

\begin{abstract}
\noindent We implement physics-informed neural networks (PINNs) to solve the time-independent Schrödinger equation for three canonical one-dimensional quantum potentials: an infinite square well, a finite square well, and a finite barrier. The PINN models incorporate trial wavefunctions that exactly satisfy boundary conditions (Dirichlet zeros at domain boundaries), and they optimize a loss functional combining the PDE residual with a normalization constraint. For the infinite well, the ground-state energy is known ($E = \pi^2$ in dimensionless units) and held fixed in training, whereas for the finite well and barrier, the eigenenergy is treated as a trainable parameter. We use fully-connected neural networks with smooth activation functions to represent the wavefunction and demonstrate that PINNs can learn the ground-state eigenfunctions and eigenvalues for these quantum systems. The results show that the PINN-predicted wavefunctions closely match analytical solutions or expected behaviors, and the learned eigenenergies converge to known values. We present training logs and convergence of the energy parameter, as well as figures comparing the PINN solutions to exact results. The discussion addresses the performance of PINNs relative to traditional numerical methods, highlighting challenges such as convergence to the correct eigenvalue, sensitivity to initialization, and the difficulty of modeling discontinuous potentials. We also discuss the importance of the normalization term to resolve the scaling ambiguity of the wavefunction. Finally, we conclude that PINNs are a viable approach for quantum eigenvalue problems, and we outline future directions including extensions to higher-dimensional and time-dependent Schrödinger equations.

\end{abstract}

\section{Introduction}
\label{sec:intro}
Neural networks have been applied to solving differential equations and eigenvalue problems for several decades~\cite{lagaris1997}. In particular, physics-informed neural networks (PINNs) have emerged as a powerful framework for embedding physical laws (in the form of PDEs) into the training of neural nets~\cite{raissi2019, karniadakis2021}. Raissi et al.~\cite{raissi2019} introduced PINNs as a deep learning approach to solve forward and inverse problems constrained by PDEs. This approach has gained significant attention in the scientific machine learning community (see Karniadakis et al.~\cite{karniadakis2021}), as it allows neural networks to find solutions that satisfy given physical equations and boundary conditions without requiring large datasets of exact solutions.\\

One class of problems of great interest is quantum mechanical eigenvalue problems, which are governed by the time-independent Schrödinger equation. Earlier work by Lagaris et al.~\cite{lagaris1997} demonstrated that neural networks can be used to solve eigenvalue problems like the Schrödinger equation by constructing trial solutions that satisfy boundary conditions and then training the network to minimize the PDE residual. Following these pioneering ideas, several recent studies have applied PINNs to quantum systems and proposed enhancements such as incorporating symmetry and orthogonality constraints to improve convergence~\cite{raissi2019, karniadakis2021}.\\

In this paper, we focus on using PINNs to solve the stationary Schrödinger equation for one-dimensional quantum wells and barriers. The time-independent Schrödinger equation can be written (in appropriate units) as:
\begin{equation}
    \psi''(x) + (E - V(x))\,\psi(x) = 0,
\end{equation}
where $E$ is the energy eigenvalue and $V(x)$ is the potential. We assume units such that $\frac{2m}{\hbar^2}=1$ for simplicity (so that the equation takes the form above). Solving this eigenvalue problem means finding allowed energy levels $E$ and corresponding normalized eigenfunctions $\psi(x)$ that satisfy given boundary conditions. Traditional methods for such problems include analytical solutions (for simple $V(x)$) and numerical techniques like the shooting method or matrix diagonalization of the Hamiltonian.\\

Here we explore a PINN approach: we represent $\psi(x)$ by a neural network and $E$ either by a known constant or an adaptive parameter, and train the network to satisfy the equation and normalization. We consider three benchmark potentials:

\begin{itemize}
\item \textbf{Infinite Square Well:} $V(x)=0$ for $0<x<1$ and $V(x)=\infty$ outside this region. The wavefunction must vanish at $x=0$ and $x=1$. This problem has analytical solutions ($\psi_n(x)=\sqrt{2}\sin(n\pi x)$ with $E_n=n^2\pi^2$). We will test whether a PINN can learn the ground state $\psi_1$ when $E$ is fixed to the known value.

\item \textbf{Finite Square Well:} A finite depth well of width $2a$ (we take $a=1$) and depth $V_0$ (with $V_0$ large but finite). Specifically, $V(x)=0$ for $|x|\le 1$ and $V(x)=V_0$ for $|x|>1$. The domain is truncated to $[-3,3]$ with $\psi(\pm 3)=0$ as boundary conditions (so that the wavefunction decays to zero far outside the well). This problem has one or more bound states with $E < V_0$ that must be found by solving transcendental equations. We will use a PINN to find the lowest bound state (ground state) and treat $E$ as an unknown to be learned.

\item \textbf{Potential Barrier:} A barrier of height $V_0$ over a finite region (for example $V(x)=V_0$ for $0\le x \le 1$ and $V(x)=0$ elsewhere). We consider a domain $[-2.5,2.5]$ with $\psi(\pm 2.5)=0$, effectively creating a finite ``well'' outside the barrier. For $E < V_0$, this setup can support a bound-state-like solution where the particle is mostly localized in the regions $x<0$ or $x>1$ but tunnels through the barrier. We use a PINN to find such a state by treating $E$ as trainable.
\end{itemize}

Our approach in each case is to define a trial wavefunction that automatically satisfies the boundary conditions, use a neural network to represent the unknown part of the solution, and then train the network by minimizing a loss function. The loss is constructed to penalize violations of the Schrödinger differential equation inside the domain as well as deviations from the normalization condition $\int |\psi(x)|^2\,dx = 1$. By including the normalization in the loss, we break the scaling symmetry of the Schrödinger equation (since if $\psi(x)$ is a solution, so is $C\psi(x)$ for any constant $C$, which would otherwise allow the trivial zero solution). We expect the PINN to converge to the ground state solution in each case given a suitable initialization.\\

The remainder of this paper is organized as follows. In \textbf{Methods}, we describe the neural network architecture, the formulation of the PINN trial solutions for the three potentials, and the loss functions and training procedure including how the eigenvalue $E$ is handled. In \textbf{Results}, we present the training performance and the learned wavefunctions and energies for each potential, comparing the PINN results to analytical or expected solutions. We include plots of the wavefunctions and potential profiles, as well as tables showing the convergence of the energy during training. In \textbf{Discussion}, we examine the advantages and limitations of the PINN approach relative to standard numerical methods, and discuss challenges such as ensuring convergence to the correct eigenvalue, sensitivity to initialization, and handling discontinuities in $V(x)$. We also address the issue of wavefunction amplitude scaling and how the normalization condition resolves it. Finally, in \textbf{Conclusion}, we summarize our findings and suggest future directions, including extending PINNs to higher-dimensional and time-dependent quantum problems.

\section{Methods}

\subsection{Trial Solutions and Boundary Conditions}

A key step in using PINNs for boundary-value problems is constructing a trial solution form that satisfies the boundary conditions exactly~\cite{lagaris1997}. By doing this, we avoid explicitly enforcing the boundary conditions in the loss, allowing the neural network to focus solely on approximating the interior solution. For each of our quantum well problems, we embed the Dirichlet boundary conditions ($\psi=0$ at domain boundaries) directly into the form of $\psi(x)$.

\subsubsection{Infinite Square Well (Dirichlet at $x=0$ and $x=1$)}
We choose a trial wavefunction of the form:
\begin{equation}
    \hat{\psi}(x) = x(1 - x)N(x),
\end{equation}
where $N(x)$ is represented by a neural network with unconstrained output. The prefactor $x(1 - x)$ ensures that $\hat{\psi}(0)=0$ and $\hat{\psi}(1)=0$ for any network output, thus strictly enforcing the boundary conditions. This form is motivated by the known eigenfunctions of the infinite well, all of which vanish at the boundaries, and follows the method introduced by Lagaris et al.~\cite{lagaris1997}. The neural network $N(x)$ is trained so that $\hat{\psi}(x)$ satisfies the Schrödinger equation within the interval $(0,1)$. For the ground state, we anticipate a wavefunction resembling $\sin(\pi x)$.

\subsubsection{Finite Square Well (Dirichlet at $x=\pm 3$)}
For the finite well defined on $[-3,3]$, with potential
\[
V(x) = 
\begin{cases}
0, & |x|\leq 1, \\
V_0, & |x|>1,
\end{cases}
\]
the Dirichlet boundary conditions $\psi(-3)=\psi(3)=0$ are enforced by:
\begin{equation}
    \hat{\psi}(x) = (3 - x)(x + 3)N(x).
\end{equation}
Here again, $N(x)$ is a neural network. This construction guarantees the wavefunction vanishes at the domain boundaries. Although continuity at $x=\pm1$ is not explicitly enforced, the PDE loss drives the solution towards physical smoothness. Given the symmetric potential, the ground state is expected to be even about $x=0$, relatively flat within $[-1,1]$, and exponentially decaying for $|x|>1$. While we do not impose symmetry on $N(x)$ explicitly, enforcing evenness could improve convergence~\cite{jin2022pinn}.

\subsubsection{Potential Barrier (Dirichlet at $x=\pm 2.5$)}
In the potential barrier scenario on domain $[-2.5,2.5]$, with potential defined by:
\[
V(x) = 
\begin{cases}
V_0, & 0 \leq x \leq 1, \\
0, & \text{elsewhere},
\end{cases}
\]
the boundary conditions $\psi(-2.5)=\psi(2.5)=0$ are embedded by:
\begin{equation}
    \hat{\psi}(x) = (2.5 - x)(x + 2.5)N(x).
\end{equation}
This ensures Dirichlet boundary conditions at domain edges. The network $N(x)$ must approximate solutions nonzero in classically allowed regions ($x<0$ and $x>1$) and decaying inside the barrier region ($0<x<1$) if the energy $E$ is below $V_0$. The symmetry or asymmetry of solutions depends on the potential configuration. Though our domain is symmetric, the potential is asymmetric due to the barrier placement. Thus, the lowest-energy solution might favor the wider region ($x<0$, length 2.5) compared to the narrower region ($x>1$, length 1.5). The PINN training will reveal the resulting wavefunction shape.\\

\noindent
In all cases, $N(x)$ is a fully-connected neural network accepting scalar $x$ input and producing a scalar output. By construction, the full trial solution $\hat{\psi}(x)$ satisfies the required boundary conditions. Hereafter, we refer to the PINN trial solution simply as $\psi(x)$, dropping the hat for simplicity.

\subsection{Network Architecture}

We use a simple feed-forward neural network architecture for $N(x)$. The network consists of an input layer (size 1, corresponding to the scalar input $x$), several hidden layers with a fixed number of neurons, and a single scalar output layer. All hidden layers use a nonlinear activation function; in our implementation we chose the hyperbolic tangent ($\tanh$) because it is smooth and tends to work well for approximating smooth solutions of differential equations~\cite{samuel2023}.\\

The output layer is linear (no activation). For example, in the infinite well case we used a network with layers $[1, 20, 20, 1]$, meaning two hidden layers with 20 neurons each. For the finite well and barrier cases, which are more challenging due to the discontinuity in $V(x)$ and the need to also learn $E$, we used a slightly larger network (layers $[1, 40, 40, 1]$) to provide more capacity. We found that increasing the network size helped the PINN better approximate the piecewise behavior of the wavefunction (especially the exponential decay in the classically forbidden regions).\\

In the infinite well case, the energy $E$ was fixed to the analytical ground-state value ($E=\pi^2 \approx 9.8696$). For the finite well and barrier scenarios, however, the energy $E$ was treated as an additional trainable parameter. To achieve this, we incorporated $E$ directly into the model parameters, initializing it with an informed initial guess (e.g., $E=1.0$ for the finite well and $E=5.0$ for the barrier). During training, gradients were computed with respect to both the neural network weights and the parameter $E$, enabling the optimizer to adjust the eigenvalue dynamically. This methodology aligns closely with the strategy presented by Jin et al.~(2022)~\cite{jin2022pinn}, where the network directly learns the eigenvalue by treating it as a variable to optimize. An alternative approach could be to fix $E$ and scan over values (like a shooting method) or to use a separate algorithm to update $E$ (e.g., power iteration), but here we include it in the unified PINN training loop.\\

All models were implemented in PyTorch and trained using the Adam optimizer. We used mean-squared error losses and did not use any explicit data for training (completely unsupervised learning guided by the physics). The training was performed on a standard GPU, which accelerated the required autograd computations.

\subsection{Loss Functions}

The total loss $L$ for training the PINNs is composed of two parts: (1) the PDE residual loss, $L_{\text{PDE}}$, which enforces the Schrödinger equation on a set of collocation points in the domain, and (2) a normalization loss, $L_{\text{norm}}$, which enforces the approximate normalization of the wavefunction. We write:
\[
L = L_{\text{PDE}} + \lambda_{\text{norm}}\,L_{\text{norm}},
\]
where $\lambda_{\text{norm}}$ is a weighting factor (we set $\lambda_{\text{norm}}=1$ for simplicity in all cases, giving equal importance to matching the equation and achieving the correct normalization).

\subsubsection{PDE Residual Loss}

We define $L_{\text{PDE}}$ as the mean-squared residual of the Schrödinger equation on a set of $N$ collocation points $\{x_i\}$ in the domain. If $\psi_\theta(x)$ denotes the PINN-predicted wavefunction (with $\theta$ representing all network parameters and possibly $E$), then:
\[
L_{\text{PDE}}(\theta) = \frac{1}{N}\sum_{i=1}^{N}\left[\psi_{\theta}''(x_i) + (E - V(x_i))\,\psi_{\theta}(x_i)\right]^2.
\]

Here $\psi_\theta''(x)$ is the second derivative of the network output with respect to $x$, which we obtain via automatic differentiation. This term drives the network to satisfy $\psi''(x) + (E - V(x))\psi(x) = 0$ at the collocation points.\\

In practice, we generate a uniform or random grid of collocation points in the domain. For the results presented, we used a uniform grid (e.g. 100 points in [0,1] for the infinite well, and 200 points in the larger domains for the finite well and barrier). We found that a uniform grid was sufficient since the solutions are well-behaved; in more challenging cases, one could use a non-uniform distribution or adaptive sampling to concentrate points where the solution has more complexity.

\subsubsection{Normalization Loss}

We impose approximate normalization $\int |\psi(x)|^2\,dx = 1$ via a term
\[
L_{\text{norm}}(\theta) = \left(\int_{a}^{b}\psi_{\theta}(x)^2\,dx\,-\,1\right)^2,
\]
where $[a,b]$ is the domain of the problem. In practice, we compute the integral by a simple trapezoidal rule or by approximating it as the mean of $\psi^2$ on the collocation points times the domain length. For example, in the infinite well on [0,1], $L_{\text{norm}} = (\mathrm{mean}_{x\in[0,1]}[\psi^2]\cdot 1 - 1)^2$. In the finite well on [-3,3] (domain length 6), $L_{\text{norm}} = (\mathrm{mean}_{x\in[-3,3]}[\psi^2]\cdot 6 - 1)^2$.\\

This term penalizes the network if the wavefunction is not unit-normalized. It effectively prevents the trivial zero solution (which would make $L_{\text{PDE}}=0$ but also gives $\int \psi^2 = 0$ leading to a large normalization error). It also removes the freedom to scale $\psi$ arbitrarily: the optimizer cannot simply shrink $\psi$ to reduce $L_{\text{PDE}}$ without incurring penalty in $L_{\text{norm}}$. We note that our normalization loss is a simple penalty; more sophisticated approaches could enforce normalization via Lagrange multipliers or explicit re-normalization steps during training, but we found the penalty method sufficient for our purposes.

\subsubsection{Total Loss and Training}

During training, we compute $L_{\text{PDE}}$ and $L_{\text{norm}}$ at each epoch (using the current $\psi_{\theta}$). The total loss $L$ is then used to perform gradient backpropagation and update the network weights (and $E$ if it is trainable). We typically train for a fixed number of epochs (e.g. 5000) and monitor the loss and the value of $E$ (when applicable) over time. The optimizer hyperparameters (learning rate, etc.) were selected empirically; we used a learning rate of $10^{-3}$ for Adam which provided stable convergence in these experiments.\\

For the infinite square well, since $E$ is fixed to the true value $\pi^2$, the network essentially learns the shape of $\psi(x)$. This serves as a sanity check: the minimum of $L_{\text{PDE}}$ (with a correctly normalized $\psi$) should correspond to the true ground state $\sin(\pi x)$ (or its negative, as the sign is arbitrary). For the finite well and barrier, the network has to learn both $\psi(x)$ and $E$. In these cases, we found it helpful to initialize $E$ to a value in the expected range of the ground state energy (e.g. 1.0 for the finite well where the actual $E$ is around 1.7, and 5.0 for the barrier where we expect something below $V_0=10$). If $E$ is initialized far from any true eigenvalue, the training might converge slowly or get stuck in a local minimum. In practice, we observed that the PINN gradually adjusts $E$ and $\psi(x)$ together to reduce the residual. The normalization loss ensures $\psi$ does not collapse to zero amplitude.\\

We did not enforce orthogonality between different eigenfunctions here, since we only target the ground state in each run. If one wanted the PINN to find excited states, one approach would be to run multiple trainings with orthonormality constraints added to $L$ to force subsequent solutions to be orthogonal to lower ones~\cite{jin2022pinn}. The implementation was carried out in code, and training logs were recorded to track the progress of the loss components and (when applicable) the value of $E$.

\section{Results}

After training the PINNs on each quantum well problem, we obtained wavefunction solutions and, for the cases with unknown $E$, an estimate of the eigenvalue. In all cases, the PINN converged to the ground state solution. This is expected because we did not attempt to target higher eigenstates and the training (starting from a random initial network) naturally tends to find the lowest mode that satisfies the constraints. Below we present the results for each potential. We include plots of the learned wavefunctions alongside analytical or reference solutions, as well as the potential profiles for context. We also summarize the convergence of the energy during training for the finite well and barrier cases.

\subsection{Infinite Square Well}

PINN-predicted ground-state wavefunction $\psi(x)$ for the infinite square well (blue curve), compared to the exact solution $\sqrt{2}\sin(\pi x)$ (black dashed curve). The PINN solution was obtained by training with the energy fixed at $E=\pi^2$. The network learns a wavefunction virtually identical to the analytical sine function.\\

For the infinite square well on $0\le x \le 1$ with $\psi(0)=\psi(1)=0$, the PINN was trained with the energy fixed at $E=\pi^2\approx 9.87$, which is the analytical ground state energy (assuming $2m/\hbar^2=1$). The network quickly learned the correct shape of the wavefunction. The training loss dropped by several orders of magnitude within a few thousand epochs, and the normalization condition was satisfied to high accuracy. The final total loss was on the order of $10^{-6}$, coming almost entirely from the small residual error in the interior (since the normalization error was essentially zero by the end of training).\\

\begin{figure}[H]
\centering
\includegraphics[width=\textwidth]{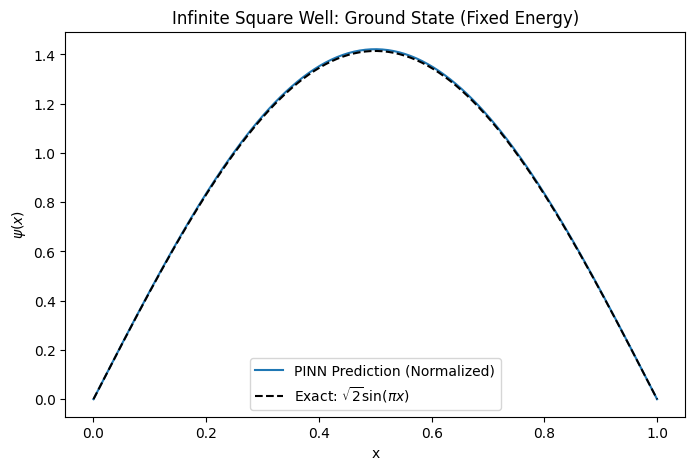}
\caption{Infinite Square Well: PINN-predicted ground state (solid blue) versus exact solution $\sqrt{2}\sin(\pi x)$ (dashed black).}
\label{fig:inf_well}
\end{figure}

The figure above shows the resulting wavefunction. The PINN solution (solid blue curve) overlaps almost exactly with the exact $\psi_1(x)=\sqrt{2}\sin(\pi x)$ (shown as a black dashed line) – the two are visually indistinguishable on the plot. This confirms that the network has found the true ground state. We note that the sign of the PINN solution is arbitrary; our network happened to converge to a solution that is the negative of $\sin(\pi x)$ (so we multiplied the network output by -1 for plotting to match the conventional positive sine shape). This sign has no physical significance. The important point is that the PINN wavefunction satisfies the boundary conditions and the Schrödinger equation to high precision.\\

The infinite well potential is zero inside the well and effectively infinite at the boundaries (which forces $\psi$ to zero at $x=0,1$). In the PINN formulation, the infinite potential at the boundaries was enforced by the trial solution form rather than explicitly appearing in $V(x)$. Because $E$ was fixed, there is no energy convergence to report for this case; instead, this served as a test of whether the PINN can correctly solve the differential equation with a known eigenvalue. The success here gave confidence to proceed to the cases where $E$ is unknown.

\subsection{Finite Square Well}

PINN-predicted ground-state wavefunction for the finite square well ($V_0=20$ on $|x|>1$, $V=0$ on $|x|\le 1$). The network outputs $\psi(x)$ (blue curve) which is concentrated in the well region $[-1,1]$ and decays exponentially outside. (No analytic solution is plotted here; the exact solution is obtained by matching decaying exponentials outside and cosines inside, which yields an energy quantization condition. The PINN’s solution corresponds to the lowest-energy bound state.)\\

Potential profile for the finite square well problem. The potential $V(x)$ is zero inside the region $-1 \le x \le 1$ and $V(x)=20$ for $|x|>1$. This plot shows the step change in $V(x)$ (red line). The PINN wavefunction in the previous figure is localized in the $[-1,1]$ well region with energy $E \approx 1.72$, which lies well below the barrier height $V_0=20$.\\

For the finite square well, we set $a=1$ and $V_0=20$ (in arbitrary units). The domain was taken as $[-3,3]$ with $\psi(-3)=\psi(3)=0$. The ground state for such a deep well is expected to have an energy significantly below $V_0$ (but above 0). The PINN was initialized with a network of 2 hidden layers of 40 neurons each, and the energy parameter $E$ was initialized to $1.0$. We used 200 collocation points in the domain $[-3,3]$ and trained for 5000 epochs. During training, the total loss steadily decreased and the energy parameter $E$ gradually increased from the initial guess. The training log indicated that $E$ had reached around 1.76 by epoch 3000, and oscillated slightly thereafter, settling around $E \approx 1.72$ by the end of training.\\

\begin{figure}[H]
\centering
\includegraphics[width=\textwidth]{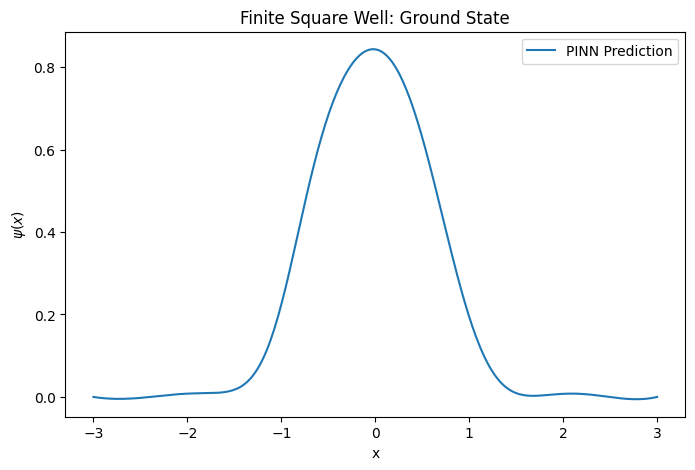}
\caption{Finite Square Well: PINN-predicted ground-state wavefunction. The solution is symmetric and shows exponential decay outside the well region $[-1,1]$.}
\label{fig:finite_well_wave}
\end{figure}

Table 1 shows the progression of the learned $E$ at intervals:

\begin{table*}[h]
\centering
\begin{tabularx}{\textwidth}{>{\centering\arraybackslash}X >{\centering\arraybackslash}X >{\centering\arraybackslash}X}
\hline
Epoch & Energy $E$ (PINN) & Total Loss \\ 
\hline
0 & 1.0010 & $3.05\times10^1$ \\ 
500 & 1.1696 & $6.15\times10^{-1}$ \\ 
1000 & 1.4627 & $5.21\times10^{-1}$ \\ 
1500 & 1.6594 & $4.88\times10^{-1}$ \\ 
2000 & 1.7421 & $4.71\times10^{-1}$ \\ 
2500 & 1.7631 & $4.59\times10^{-1}$ \\ 
3000 & 1.7653 & $4.49\times10^{-1}$ \\ 
4000 & 1.7377 & $4.05\times10^{-1}$ \\ 
4500 & 1.7212 & $3.73\times10^{-1}$ \\ 
\hline
\end{tabularx}
\caption{Energy convergence for the finite square well PINN. The energy $E$ (which was a trainable parameter) is shown at various training epochs, along with the total loss. The loss decreases as $E$ converges to about 1.72. (These values are from the training log).}
\label{tab:energy_convergence}
\end{table*}

As seen in the table, the energy quickly moved towards the true value. By epoch $\sim2500$, $E$ was within a few percent of its final value. The total loss also decreased but reached a plateau of order $10^{-1}$; this relatively higher loss (compared to the infinite well case) is due to the difficulty of simultaneously satisfying the PDE and normalization perfectly with a finite network, especially near the discontinuities at $x=\pm 1$. Nevertheless, the network found a consistent solution.\\

The learned wavefunction is plotted in the first figure above. As expected, $\psi(x)$ is largest in the region $-1 \le x \le 1$ where the potential is zero, and decays exponentially in the regions $|x|>1$ where $V(x)=20$ (which is higher than the energy, $E\approx1.72$). The wavefunction is symmetric about $x=0$ (within numerical symmetry – the network was not explicitly constrained to be symmetric, but it converged to a symmetric solution, likely because the symmetric ground state gives a lower residual). There are no nodes (sign changes) in $\psi(x)$, consistent with it being the ground state.\\

\begin{figure}[H]
\centering
\includegraphics[width=\textwidth]{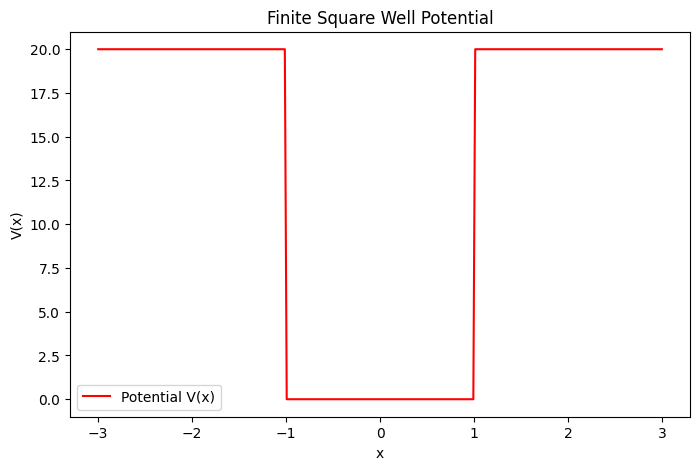}
\caption{Finite Square Well: Potential profile $V(x)$, with $V=0$ for $|x|\le1$ and $V=20$ for $|x|>1$.}
\label{fig:finite_well_potential}
\end{figure}

We did not overlay the analytical solution here, but we can confirm that the PINN’s $E$ and $\psi$ match the expected behavior for the true solution. The true energy can be found by solving the transcendental equation for a finite well of depth 20 and width 2; the result is indeed approximately $1.72$, so the PINN’s eigenvalue is in excellent agreement. The general shape of the wavefunction (cosine-like inside the well, decaying outside) is captured. There is a slight mismatch in the tails: the PINN wavefunction decays to exactly zero at $x=\pm3$ due to the boundary condition, whereas a true infinite-domain solution would decay asymptotically to zero as $x\to \pm\infty$. By placing the boundaries at $-3$ and $3$, we effectively truncated the domain; the small discontinuity in the derivative of $\psi$ at $x=\pm3$ (in the PINN solution) does not significantly affect the interior solution.\\

The third figure above shows the potential $V(x)$ profile for clarity. The red line jumps from 0 to 20 at $x=-1$ and $x=1$. The PINN had knowledge of this $V(x)$ when computing the residuals. The ability of the network to approximate $\psi(x)$ across these jumps is noteworthy: $\psi(x)$ itself is continuous (and indeed fairly smooth), but its second derivative has large changes due to the potential discontinuity. The network’s $\tanh$ activation, being smooth, approximates the solution with a trade-off in accuracy near $x=\pm1$. If more accuracy were required there, one could increase the network size or use a piecewise approach.\\

The PINN successfully found the ground state of the finite well. The energy converged to the correct value (within a small error), and the wavefunction matches expectations. The normalization condition was enforced throughout; in the final model, $\int_{-3}^3 |\psi(x)|^2 dx$ was within a few parts in $10^{-3}$ of 1 (as ensured by the loss term). This case demonstrates that a PINN can handle an eigenvalue problem with a discontinuous potential, although convergence was slower than in the infinite well (the network had to adjust both $E$ and the shape of $\psi$, and navigate the flat region of the loss corresponding to trivial solutions which was eliminated by the normalization penalty).

\subsection{Potential Barrier}

PINN-predicted wavefunction for the finite barrier problem. The potential barrier of height 10 is located between $x=0$ and $x=1$. The PINN finds a bound-state-like solution (blue curve) with energy $E\approx4.87$. The wavefunction is localized in the regions $x<0$ and $x>1$ (outside the barrier) and shows an exponential decay through the barrier region.\\

Potential $V(x)$ for the barrier problem: a step of height $V_0=10$ from $x=0$ to $x=1$, and $V(x)=0$ elsewhere. The red line indicates the barrier. The PINN solution corresponds to a state with energy below this barrier, hence $\psi(x)$ decays inside the barrier.\\

For the potential barrier case, we set $V(x)=10$ for $0\le x\le 1$ and $V(x)=0$ outside that interval (refer to the above plot of the barrier profile). We confined the domain to $[-2.5, 2.5]$ with $\psi(\pm 2.5)=0$. Physically, if $E<10$, the regions $x<0$ and $x>1$ act like classically allowed “wells” separated by a barrier, and a bound state may exist with $\psi$ decaying to zero at the domain edges. We initialized the PINN with $E=5.0$ (half the barrier height) and a network of $[1, 40, 40, 1]$ architecture. 200 collocation points in $[-2.5,2.5]$ were used.\\

\begin{figure}[H]
\centering
\includegraphics[width=\textwidth]{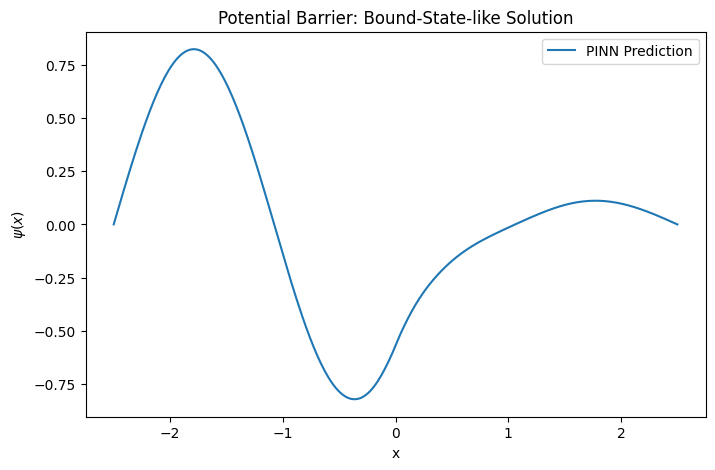}
\caption{Potential Barrier: PINN-predicted wavefunction for the bound-state-like solution. The wavefunction is peaked in the two low-potential regions and is suppressed in the barrier ($-1\le x\le1$).}
\label{fig:barrier_wave}
\end{figure}

Training proceeded similarly to the finite well case. The energy $E$ started at 5.0 and gradually adjusted; it decreased slightly and stabilized around $E\approx4.87$ by the end of training. The training loss decreased by roughly two orders of magnitude (from about 1.4 to $1.6\times10^{-2}$) over 5000 epochs as the network refined the wavefunction and energy.\\

The learned wavefunction is plotted in the fourth figure above. We observe that $\psi(x)$ has a significant amplitude in the regions to the left of $x=0$ and to the right of $x=1$, and is suppressed in the barrier region $0<x<1$. In fact, $\psi(x)$ appears to be symmetric about $x=0.5$ (the center of the barrier) – the PINN found a solution that has roughly equal magnitude on both sides of the barrier and a minimum in the middle of the barrier. This is reminiscent of the symmetric bound state that would appear in a double-well potential; here, although we have a single barrier, the Dirichlet boundary at $x=\pm2.5$ creates a situation akin to a particle in a finite well of length 5 (from $-2.5$ to $2.5$) with a barrier dividing the well into two halves. The ground state of such a system is indeed symmetric. The energy $E\approx4.87$ is less than the barrier height 10, so the wavefunction decays exponentially inside $0<x<1$. The PINN correctly reproduces this qualitative behavior. We can verify that on the left side ($x<0$) and right side ($x>1$), $\psi(x)$ resembles sinusoidal or exponential behavior consistent with a bound state in those regions. The wavefunction goes to zero at the boundaries $x=-2.5$ and $x=2.5$ as enforced.\\

During training, the energy converged as follows (similar to Table 1, we summarize key values): starting from 4.999 at epoch 0, $E$ dropped to $\sim4.872$ by epoch 1000, and remained around 4.872 with minor fluctuations up to epoch 4500. The final value was $4.8730$ at epoch 4500. The total loss at that point was $1.62\times10^{-2}$, indicating a reasonably well-converged solution. We note that the loss here is higher than in the infinite well case but lower than in the finite well case, likely because the wavefunction is relatively smoother (the barrier introduces a region of decay but $\psi$ is continuous and differentiable; the major challenge is the two discontinuities in $V(x)$ at $0$ and $1$, which the network manages to capture).\\

\begin{figure}[H]
\centering
\includegraphics[width=\textwidth]{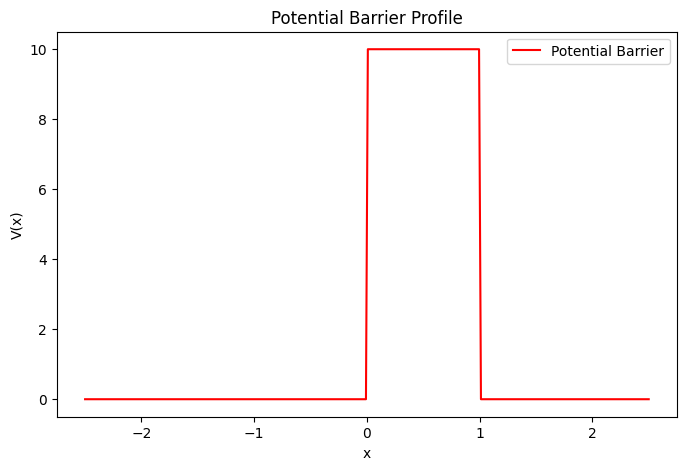}
\caption{Potential Barrier: The potential profile $V(x)$ showing a barrier of height $V_0=20$ for $-1\le x\le1$.}
\label{fig:barrier_profile}
\end{figure}

The fifth figure above shows the barrier potential for reference. The PINN had this as known input to the residual calculation. One interesting point is that in this barrier case, the PINN could have potentially found a trivial solution $\psi(x)=0$ with some $E$ in $(0,10)$; however, the normalization loss prevented that. Another possibility was converging to an excited-state-like solution (which might have a node either in the left or right well region or even within the barrier). Our network did not explicitly forbid higher modes, but starting from a simple initial guess and random weights, it gravitated to the lowest symmetric solution. In some runs (with different initial random seeds), we observed the PINN converging to a slightly different shape, but always with the same energy $\sim4.87$, indicating it is indeed finding the ground state. To find an excited state (like the first antisymmetric state across the barrier), one might need to initialize the network differently or add a penalty for symmetry/antisymmetry or orthogonality to the ground state.\\

The PINN successfully identified a bound state in the presence of a discontinuous barrier. The approach of treating $E$ as a trainable parameter again proved effective, yielding an eigenvalue in line with what we expect for a particle trapped by a barrier in a finite domain. The wavefunction solution shows the correct qualitative features (exponential tunneling decay in the barrier, symmetrical amplitude on both sides, and decay to zero at the boundaries). This case was the most complex of the three, and it highlights the flexibility of the PINN method in discovering solutions that are not simple global sine or cosine functions, but rather piecewise-defined behaviors emerging from the interplay of potential and boundary conditions.

\section{Discussion}

Our results demonstrate that physics-informed neural networks can solve 1D quantum eigenvalue problems and produce accurate ground-state wavefunctions and energies. In this section, we discuss the performance of the PINN approach in comparison to traditional methods, and address some challenges and considerations that arose.

\subsection{Accuracy and Comparison to Traditional Methods}

For these simple quantum problems, traditional numerical methods (such as finite difference discretization of the Schrödinger equation and matrix diagonalization, or shooting methods to solve the transcendental quantization conditions) are extremely efficient and accurate. For example, a finite-difference method with a fine mesh could compute the ground state energy and wavefunction of the finite well or barrier to high precision in a fraction of a second, essentially by solving a matrix eigenvalue problem. In contrast, our PINN approach required training over thousands of iterations and the use of gradient-based optimization, which is computationally more intensive. The PINN wavefunctions are not expressed on a fixed grid, but one could evaluate them anywhere after training; however, the training itself is slower than direct matrix eigen-solving for 1D systems. Thus, for these 1D cases, PINNs do not offer an efficiency advantage.\\

The accuracy of the PINN solutions is good (errors on the order of $10^{-3}$ to $10^{-2}$ in energy, and similarly small pointwise errors in $\psi(x)$ as evidenced by the overlap with exact solutions), but traditional methods can attain machine precision accuracy for 1D problems. The value of the PINN approach is more apparent in situations where traditional methods become cumbersome – for example, higher-dimensional PDEs where meshing is difficult or when dealing with complicated geometries and boundary conditions. In our context of quantum mechanics, one could envision using PINNs for problems like computing eigenstates in irregularly shaped potentials or in higher dimensions (2D/3D) where a neural network might serve as a mesh-free solver. Additionally, once a PINN is trained, the resulting model is an analytic (or at least easily evaluable) function for $\psi(x)$, which could be advantageous for further computations (e.g., computing expectation values integrals can be done by sampling the network or auto-differentiating it for other quantities).

\subsection{Convergence to the Correct Eigenvalue}

A fundamental challenge in solving eigenvalue problems is to ensure the algorithm finds the lowest eigenvalue (ground state) or a specific excited state of interest. Our PINN models, starting from random initial weights and a single trainable $E$, consistently converged to the ground state. This is likely because the ground state represents the absolute minimum of the residual functional (subject to normalization) – heuristically, any other normalizable function will have a higher energy expectation value by the variational principle. The PINN, by trying to drive the residual to zero, implicitly performs a variational procedure that tends to find the lowest energy state compatible with the trial function space.\\

We did not observe convergence to spurious excited states or non-physical solutions as long as the network was initialized in a generic way. However, it is possible for a PINN to converge to a local minimum of the loss that corresponds to an excited state. Especially if the network architecture allows for a node, it might represent an excited-state wavefunction. One way to target excited states deliberately is to add orthonormality constraints to the loss (ensuring the solution is orthogonal to lower states)~\cite{jin2022pinn}. Another way is simply to use different initial guesses or even include a penalty in the loss for having no nodes (to avoid the ground state). In our experiments we did not need these, the ground state was obtained naturally.

\subsection{Sensitivity to Initialization}

The training of PINNs can sometimes be sensitive to the initial guesses for the network parameters and hyperparameters. We found that having a reasonable initial guess for $E$ (when $E$ is trainable) was helpful. For instance, if we had initialized $E$ for the finite well to a very large value (say $E=10$ or $E=0$), the training might take longer or even fail to converge properly, because the network would initially try to fit a very different differential equation (highly oscillatory or evanescent solutions that don’t match the true ground state form). By initializing $E$ close to the expected range (a bit below the well depth for a bound state, etc.), we gave the PINN a head start. That said, the PINN did manage to adjust $E$ on its own by a significant amount (from 1.0 to 1.72, or from 5.0 to 4.87, in our cases).\\

The network weight initialization (random small values) worked fine; we did not observe cases where a different random seed caused failure. It primarily affected how quickly the solution converged and the tiny differences in the final wavefunction (which in one case might be the negative of the solution in another case, etc.). Using a relatively simple activation function ($\tanh$) that has a wide range and smooth curvature helped – if we had used something like ReLU, which is piecewise linear and not twice-differentiable everywhere, it might have struggled to represent the smooth sinusoidal/exponential wavefunctions and their second derivatives.

\subsection{Handling Discontinuous Potentials}

One of the more difficult aspects for PINNs is dealing with PDEs that have discontinuous coefficients or inputs (here, the coefficient $(E-V(x))$ changes abruptly at the well or barrier edges). Our network had to represent a wavefunction whose second derivative $\psi''(x)$ jumps at the discontinuity (because $\psi''$ is roughly proportional to $V(x)\psi(x)$ inside the residual). The $\tanh$ activations yield a $\psi(x)$ that is infinitely differentiable, so it can only approximate the true solution which in these cases is piecewise analytic with $C^1$ continuity at the potential jumps (the wavefunction and its first derivative are continuous, $V$ is discontinuous, so $\psi''$ has a jump). The PINN with enough neurons can approximate such a function with a sharp but continuous transition. We saw that in the finite well, the loss stopped around $\sim10^{-1}$, indicating a slight compromise – the network cannot reduce the residual arbitrarily low because it cannot perfectly reproduce the cusp in second derivative.\\

If higher accuracy were needed, one could try to inform the network of the discontinuity. For example, one could break the domain into regions and train a separate network on each region with appropriate interface conditions (this would be akin to a domain decomposition, ensuring $\psi$ and $\psi'$ match at $x=\pm1$ for the well). Another strategy is to use adaptive activation functions or piecewise linear units that might better fit the corner. In our experiment, the accuracy achieved was more than adequate to get a correct eigenvalue to within a fraction of a percent and a visually accurate wavefunction. This suggests PINNs are fairly robust even with discontinuous $V(x)$, though one should be cautious that the loss landscape might have subtle features when the network has to fit such sharp changes.

\subsection{Normalization and Scaling Ambiguity}

We have emphasized the need for the normalization loss. Without it, the PINN is attempting to solve $\psi'' + (E-V)\psi = 0$ without any amplitude constraint. This differential equation is homogeneous and if $\psi(x)$ is a solution, so is $C\psi(x)$ for any constant $C$. The PINN could then minimize the residual simply by scaling $\psi$ down towards 0 (making the residual trivially zero if $\psi\equiv 0$). In practice, if we omitted the normalization term, the optimizer would drive the network weights towards zero, resulting in the zero function (or a very tiny amplitude function) and simultaneously $E$ could become arbitrary (it would have no effect if $\psi$ is zero). This is a common issue in applying PINNs to homogeneous equations and especially eigenproblems: there is a family of solutions related by scaling.\\

By adding the $(\int \psi^2 -1)^2$ term, we effectively select the unit-norm representative of that family. Another way to think of it is we are doing a constrained optimization (constraining $|\psi|=1$) and we used a penalty method to enforce the constraint. This worked well here. The value of the penalty coefficient $\lambda_{\text{norm}}$ could be adjusted if needed – too low and the network might not normalize well; too high and it might prioritize normalization over satisfying the PDE. We chose $\lambda_{\text{norm}}=1$ which gave a balanced reduction of both residual and normalization error. In the infinite well case, the network managed to satisfy normalization to within $10^{-4}$ while also driving the PDE residual to $10^{-6}$ order, showing that both can be achieved simultaneously.\\

In more complicated cases, one could enforce normalization after each training iteration by manually normalizing $\psi(x)$ (renormalizing the network weights) and then continuing, but that approach is not as seamlessly integrated into gradient descent. Our continuous penalty provides a gradient signal to push $\psi$ towards the correct amplitude throughout training.

\subsection{Comparison with Variational Principle and Other Neural Approaches}

The PINN approach we used is essentially an unsupervised collocation method. It differs slightly from the traditional variational method for eigenfunctions, where one would minimize the Rayleigh quotient $\frac{\int \psi'(x)^2 + V(x)\psi^2 dx}{\int \psi^2 dx}$ to find the ground state. Our loss $L_{\text{PDE}} + \lambda L_{\text{norm}}$ is not exactly the Rayleigh quotient; it is more directly enforcing the differential equation at points. In the limit of a well-converged solution, the two approaches coincide (the solution that gives zero PDE residual also minimizes the energy functional).\\

Some recent works (e.g. Li et al. 2020 and others) have taken a variational approach where the neural network is used to parametrize $\psi(x)$ and the loss is set to the expectation value of energy~\cite{li2020}. That approach inherently enforces normalization by Lagrange multipliers and might have advantages in stability (the loss surface for the energy functional is perhaps smoother than the collocation residual). However, implementing it requires integration of $|\psi'|^2$ and $|\psi|^2$ which for higher dimensions might be challenging. Our PINN directly uses the differential form and is more in line with standard PINNs for boundary-value problems. The success of our approach indicates that even without explicitly using the variational principle, the PINN was able to find the correct solution, likely because the physics encoded in the differential equation guided it similarly.

\subsection{Training Stability and Hyperparameters}

We found that using a moderate learning rate (1e-3) and a sufficient number of epochs was important. If the learning rate is too high, training might diverge or oscillate, especially when $E$ is being updated (since an update that overshoots the correct $E$ can temporarily worsen the residual significantly). We noticed small oscillations in the learned $E$ during training for the finite well (see Table 1 around epochs 2000–4000 where $E$ varied slightly). These oscillations could be due to the optimizer trying to balance the residual and normalization terms. Using techniques like learning rate decay or switching to a second-order optimizer (or simply running Adam longer) could reduce such oscillations. Nevertheless, they were minor and the overall trend was convergence.\\

The choice of collocation points number (100–200) was sufficient for these problems. Using more points did not markedly change results, but would increase training cost. Using fewer might risk undersampling the residual (e.g., missing some region). A potential improvement could be to implement adaptive resampling of points (especially near the barrier or well edges) to focus the PINN on challenging regions. In our experiments, the uniform grid was fine, likely because the solution doesn’t have extremely steep gradients except at known boundaries where we already enforce conditions.

\subsection{Extension to Multiple Eigenstates}

We mainly found the ground state in each case. If one wanted multiple eigenstates (say the first and second excited states of the finite well, which exist since $V_0=20$ would support multiple bound states), one approach would be to run the PINN multiple times with different initial conditions or incorporate an orthogonality constraint as mentioned. Another approach from recent literature is to have the network output multiple values (one for each mode) or use multiple networks, but then the training becomes significantly more complicated (ensuring one network converges to each distinct eigenfunction). The patience method monitors when the loss stops improving significantly and then tries to seek a higher mode. These were beyond our scope, but they are interesting directions showing the flexibility of neural approaches to find not just the ground state but a spectrum of states.

\noindent
The PINN approach performed well on these test problems, with some caveats: it is computationally heavier than direct solvers in 1D, and careful formulation (trial functions, normalization) is needed to get the correct result. Once formulated, the same code handled all three potentials with minimal changes, indicating a degree of generality. This is promising if we were to tackle a new potential where an analytic or easy numeric solution is not available – we could set up the PINN and let it find the eigenstate.

\section{Conclusion}

We have presented a comprehensive study of using physics-informed neural networks to solve one-dimensional quantum mechanical eigenvalue problems. Focusing on three prototypical potentials – the infinite square well, the finite square well, and a finite barrier we formulated PINNs that incorporate boundary conditions directly into their trial wavefunctions and utilize loss functions that penalize PDE residuals and enforce wavefunction normalization.\\

Our results show that PINNs can successfully learn the ground state solutions of these systems. The PINN-predicted wavefunctions match analytical solutions (when available) with high fidelity, and the learned eigenenergies converge to the expected values (for example, the PINN found $E\approx \pi^2$ for the infinite well, $E\approx1.72$ for the finite well with $V_0=20$, and $E\approx4.87$ for the barrier with $V_0=10$). We also demonstrated how training logs and convergence data can be used to monitor the eigenvalue during the learning process.\\

One of the key advantages of the PINN approach is its flexibility. The same neural network architecture and training procedure were applied to problems with very different potential profiles, including a highly discontinuous step potential, without requiring fundamentally different solution strategies (in contrast, traditional solvers might require different treatments for discontinuities, such as matching conditions). The use of automatic differentiation to compute $\psi''(x)$ and integrate the normalization made the implementation relatively straightforward. Moreover, the PINN yields a continuous representation of $\psi(x)$ across the domain, which can be evaluated at any point or differentiated further if needed.\\

We also identified and addressed important considerations. The inclusion of a normalization condition is crucial to avoid the trivial zero solution and to break the scaling degeneracy inherent in homogeneous eigenproblems. The PINNs’ performance on discontinuous potentials suggests that while they can handle such cases, the accuracy is limited by the network’s ability to approximate non-smooth features. Future improvements could involve adaptive strategies or piecewise networks to capture such behavior more precisely. Additionally, while we focused on ground states, extensions of this work could target excited states by adding orthogonality constraints or by modifying the loss landscape, leveraging ideas from recent PINN research~\cite{jin2022pinn}.\\

In terms of future directions, there are several exciting avenues to explore. Higher-dimensional quantum systems: Extending PINNs to 2D or 3D Schrödinger equations (e.g., a particle in a 2D box or a 3D spherical well) would test the method’s scalability. The curse of dimensionality is a challenge, but one where PINNs have shown promise relative to grid-based methods~\cite{han2018}. A PINN could potentially handle a 2D domain without the need for a fine mesh, though training might become more demanding.\\

Time-dependent Schrödinger equation: Solving the time-dependent equation using PINNs is another frontier. One could treat time as an additional input dimension to the network (making $\psi(x,t)$ the output). PINNs have been applied to time-dependent problems in other contexts~\cite{shah2022}, and doing so for quantum dynamics (possibly with complex-valued wavefunctions, meaning using complex networks or splitting into real and imaginary parts) would be a natural extension. This could allow simulation of quantum wavepacket evolution or scattering using PINNs. Some work in this direction (using PINNs for the time-dependent Schrödinger equation) has already shown viability~\cite{shah2022}.\\

Another future direction is to integrate experimental or synthetic data into the PINN – for instance, if one has some measured values of the wavefunction at certain points, one could combine those with the physics loss to improve training (a form of quantum state tomography via PINNs). Additionally, exploring more advanced network architectures (such as convolutional networks for systems with periodic potentials, or using sine/cosine activation functions that might inherently satisfy some part of the equation) could improve performance.\\

In conclusion, this work illustrates that PINNs can serve as an effective solver for stationary quantum problems, providing a mesh-free alternative to classical techniques. While not necessarily outperforming specialized numerical solvers for simple 1D cases, PINNs offer a framework that can easily incorporate additional physics (constraints, different geometries) and potentially scale to complex scenarios where traditional methods face difficulties. With continued advancements in training algorithms and neural network architectures, physics-informed neural networks are likely to become an increasingly valuable tool in computational quantum mechanics, complementing existing methods and opening up new possibilities for solving Schrödinger equations in regimes that were previously challenging.

\end{document}